\documentstyle[preprint,aps,epsf]{revtex}

\begin{document}
\draft


\title{Discrete Solitons and Breathers with 
Dilute Bose-Einstein Condensates}
\author{Andrea Trombettoni and Augusto Smerzi}
\address{
 Istituto Nazionale di Fisica per la Materia and  
International School for Advanced Studies,\\
 via Beirut 2/4, I-34014, Trieste, Italy}
\date{\today}
\maketitle
\begin{abstract}
We study the dynamical phase diagram 
of a dilute Bose-Einstein condensate (BEC)
trapped in a periodic potential.
The dynamics is governed by a discrete
non-linear Schr\"odinger equation: intrinsically localized excitations, including discrete solitons
and breathers, can be created even if the BEC's interatomic potential 
is repulsive.
Furthermore, we analyze the Anderson-Kasevich experiment [Science {\bf 282}, 1686 (1998)], pointing out that mean field effects lead to a coherent destruction of the interwell Bloch oscillations. 
\end{abstract}
\pacs{PACS: 63.20.Pw, 05.45.-a}


Localization phenomena are ubiquitous in physics and in biology 
\cite{special}. 
Intrinsically localized excitations, as
solitons (shape preserving) and breathers (characterized by internal
oscillations), are important channels 
for energy transport in non-linear media,
such as optical fibres and waveguides \cite{flach98}, polaronic materials \cite{andersen93}
and biological molecules \cite{peyrard93}.
Intense theoretical research is now focusing  
on the existence of solitons and breathers in a lattice 
(often named discrete solitons and 
discrete breathers (DSB) \cite{flach98}), such as those appearing 
in quantum systems governed by a discrete
non-linear Schr\"odinger equation (DNLSE) 
\cite{hennig99,rasmussen00,kevrekidis00}.
Current approaches include the search for exact 
solutions in some limits \cite{cai94}; 
effective (point) particle and variational approaches \cite{scharf91,cooper93,aceves96}; perturbation 
around the linearized case and, of course, numerical solutions 
\cite{flach98,hennig99}. 
Although intensely studied, DSB have been 
experimentally observed only quite recently in superconducting ladders of 
Josephson junctions \cite{binder00}, in antiferromagnet 
systems \cite{schwartz99}, in optical waveguides \cite{eisenberg98}
and in low dimensional materials \cite{swanson99}.

The discrete solitons/breathers are characterized by a dynamical,
self-maintained energy localization, due to both 
the discreteness and the non-linearity of the underlying equations of 
motion. The discreteness provides a band structure of the 
excitation spectrum, while
the non-linearity allows for the tuning of the DSB energy outside
the band. The finite energy gap guarantees the (meta-)stability of the DSB's.
These, obviously, have a different nature to
the "Anderson localizations", created by impurities or imperfections 
of the lattice \cite{ramakrishnan86};
the incorporation of disorder into non-linear excitations, 
and the tracing out of its dynamical effects
are also important theoretical problems.

Bright solitons can occur in spatially homogeneous, dilute 
Bose-Einstein condensates (BEC) with an attractive interatomic interaction  
(s-wave scattering length $a < 0$) 
\cite{jaksch98,zubay99}. Dark solitons, 
propagating density dips, have been predicted and experimentally observed
in BEC's with a repulsive interaction ($a > 0$) \cite{burger99}.
The dynamics of a BEC trapped in a spatially periodic
potential \cite{berg98,javanainen99,chiofalo00}, on the other hand, can be mapped,
in the tight binding approximation, to a DNLSE.
Bright discrete solitons and breathers can be created in DNLSE 
even with a repulsive BEC interatomic interaction.

In Ref. \cite{anderson98} a one-dimensional, vertical optical array 
was created by two counterpropagating laser beams. A weakly interacting 
Bose-Einstein condensate was trapped in $\sim 30$ wells, situated at the 
antinodes of the standing optical wave. Each well contained 
approximately $1000$ condensate atoms, with the peak densities matching 
a gaussian profile. The lowest Bloch band dynamics 
of this system maps on a DNLSE, obtained by discretizing the 
Gross-Pitaevskii equation governing the condensate dynamics 
in the periodic potential. Since the array is oriented vertically, the
atoms undergo coherent Bloch oscillations, driven by the interwell 
gravitational potential. At the edge of the Brillouin zone, a fraction 
of atoms can Zener tunnel in the higher energy band which, 
in this specific case, is in the continuum. 
A coherent leakage from the trap was observed at each Bloch period.
However, an increase of the on-site particle densities washed out the coherent signal. 
This has been understood as a drift of the relative phases between wells due to mean field effects \cite{anderson98}.  

In this Letter we study the DNLSE using a variational approach: we show that 
a (complex) gaussian variational ansatz 
(describing a soliton / breather profile, combined with a Lagrangian 
optimization) yields a coupled dynamics 
of the profile parameters. A variety of discrete solitons / breathers 
are investigated analytically and compared with numerical 
calculations. This comparison is surprisingly successful in describing 
even details of the quite complex dynamical and collisional behaviour. 
Stability phase diagrams for such states are obtained
by inspection of the profile dynamics equations.

The full Bose condensate dynamics satisfies the
Gross-Pitaevskii equation (GPE) \cite{dalfovo99}:
\begin{equation}
\label{GPE}
i \hbar \frac{\partial \Phi}{\partial t}= - \frac{\hbar^2}{2 m} 
\nabla^2 \Phi + [V_{ext} + g_0 \mid \Phi \mid^2] \Phi 
\end{equation}
where $V_{ext}$ is the external potential and 
$g_0=\frac{4 \pi \hbar^2 a }{m}$,
with $m$ the atomic mass. 
For a tilted trap as in \cite{anderson98} the external potential 
is given by the sum of the 
gravitational potential and the laser field
\begin{equation}
\label{potential_anderson_kasevich}
V_{ext}(\vec{r})=m g z + U_L(x,y) \sin^2 [ 2 \pi z / \lambda ] 
\end{equation}
where $\lambda$ is the wavelength of the lasers (the spacing in the 
lattice is $\lambda / 2$) and $U_L(x,y)$ is determined by the transverse 
intensity profile of the (nearly gaussian) laser beams. In 
\cite{anderson98} $\lambda=850$ nm and the $1/e^2$ 
radius of the transverse profile is $\approx 80 \mu$m, an order of 
magnitude larger than the transverse radius of the condensate. The well 
depths scale linearly with the intensity of the beam. At the center of 
the beam the trap depths are $1.4 E_R$ where $E_R=\frac{\hbar^2 k^2}{2m}$ 
is the recoil energy ($k=\frac{2\pi}{\lambda}$).  

In the tight-binding approximation the condensate order parameter can be 
written as:  
\begin{equation}
\label{TB}
\Phi (\vec{r},t)= \sqrt{N_T} \, \sum_{n} \psi_n(t)~ \phi(\vec{r}-\vec{r}_n),
\end{equation}
where $N_T$ is the total number of atoms and $\phi(\vec{r}-\vec{r}_n) $ is the condensate wave function
localized in the trap $n$ with $\int d \vec r~ \phi_n ~\phi_{n+1} \simeq 0$,
and $\int d \vec r ~\phi_n^2 = 1$.
$\psi_n=\sqrt{\rho_n(t)} \, e^{i \theta_n (t)}$ is the $n$-th amplitude 
($\rho_n=N_n/N_T$, where $N_n$ and $\theta_n$ are the number of particles and phases in the 
trap $n$). 
Replacing the ansatz (\ref{TB}) in (\ref{GPE}) we find that the
GPE reduces to a DNLSE:
\begin{equation}
\label{DNLS}
i  \frac{\partial \psi_n}{\partial t} = - \frac{1}{2}  
(\psi_{n-1}+\psi_{n+1}) + (\epsilon_n+ \Lambda \mid \psi_n \mid ^2)\psi_n 
\end{equation}
where $\epsilon_n= \frac{1}{2K} \int d\vec{r} \, \big[ \frac{\hbar^2}{2m} 
(\vec{\nabla} \phi_n )^2+V_{ext} \phi_n^2 \big]$, 
$\Lambda = \frac{g_0 N_T}{2K} \int d\vec{r} \, 
\phi_n^4 $ and $K \simeq - \int d\vec{r} \, \big[ \frac{\hbar^2}{2m} 
\vec{\nabla} \phi_n \cdot \vec{\nabla} \phi_{n+1} + \phi_n V_{ext} \phi_{n+1}
 \big] $; the time has been rescaled as
$t \to \frac{\hbar}{2 K} t$. Eq.~(\ref{DNLS}) is the equation of motion 
$\dot{\psi_n} = \frac{\partial \cal{H}}{\partial (i \psi^\ast_n)}$,
where $\cal{H}$ is the Hamiltonian function
\begin{equation} 
\label{HAM}
{\cal{H}} = 
{\sum_n} [ -\frac{1}{2} ( \psi_n \psi^\ast_{n+1} + \psi^\ast_n
\psi_{n+1} )
+ \epsilon_n \mid
\psi_n\mid^2 + {\Lambda \over 2} \mid\psi_n\mid^4  ]
\end{equation}
with $i \psi_n^\ast, \psi_n$ canonically conjugate variables.
Both the Hamiltonian $\cal H$ and the norm 
$\sum_n \mid \psi_n \mid ^2 = 1$ 
are conserved. 

To analyze the Anderson-Kasevich experiment,
we study the dynamical evolution of 
a gaussian profile wave packet, which we parametrize as
$\psi_V^{n}(t)=\sqrt{k} \, \cdot \, \exp 
\Big \{ -\frac{(n-\xi)^2}{\gamma^2} + ip(n-\xi) + 
i \frac{\delta}{2}(n-\xi)^2 \Big \}$
where $\xi(t)$ and $\gamma(t)$ are, respectively, 
the center and the width (in the lattice units) 
of the density $\rho_n = \mid \psi_n \mid^2$, 
$p(t)$ and $\delta(t)$ 
their associated momenta and $k(\gamma,\xi)$ a normalization factor. 
The wave packet dynamical evolution can be obtained by a 
variational principle from the Lagrangian
${\cal L}= \sum_n  i \dot{\psi}_n \psi_n^\ast - \cal{H}$,
with the equations of motion for the variational parameters $q_i(t)=\xi,\gamma,p,\delta$
given by $\frac{d}{d t} \frac{\partial {\cal L}}{\partial {\dot q}_i} =
\frac{\partial {\cal L}}{\partial  q_i} $. After some algebra we obtain
\cite{nota1} ${\cal L}=p \dot{\xi} - \frac{\gamma^2 \dot{\delta}}{8} - \frac{\Lambda}
{2 \sqrt{\pi} \gamma} + \cos{p} \, \cdot \, e^{-\eta} - V(\gamma,\xi)$
where $\eta = \frac{1}{2 \gamma^2} + \frac{\gamma^2 \delta^2}{8}$ and
$V=k \int\limits_{-\infty}^{\infty} dn \, \epsilon_n e^{-\frac{2(n-\xi)^2}{\gamma^2}}$. 
The variational equations of motion become: 
\begin{equation}
\label{p-xi} 
\left\{\begin{array}{ll} 
\dot{p} = - \frac{\partial V}{\partial \xi} \\
\dot{\xi}=\sin{p} \, \cdot \, e^{-\eta}
\end{array}
\right.
\end{equation}
\begin{equation} 
\label{alpha-delta}
\left\{\begin{array}{ll} 
\dot{\delta} = \cos{p} \Big(\frac{4}{\gamma^4}-\delta^2 \Big) 
e^{-\eta} + \frac{2 \Lambda}{\sqrt{\pi} \gamma^3} - \frac{4}{\gamma} \frac{\partial V}{\partial \gamma}\\
\dot{\gamma}= \gamma \delta \cos{p} \, \cdot \, 
e^{-\eta}
\end{array}
\right.
\end{equation}
with the pairs $\xi, p$ and $\frac{\gamma^2}{8}, \delta$ which are canonically conjugate 
dynamical variables with respect to the effective Hamiltonian
\begin{equation}
\label{Hamiltonian-DNLS}
H = \frac{\Lambda}{2\sqrt{\pi} \gamma} -  \cos{p} \, \cdot \, 
e^{-\eta} + V(\gamma, \xi).
\end{equation} 

The wave packet group velocity is given by  
$v_g \equiv \frac{\partial H}{\partial p} = \dot{\xi} = \tan{p} / {m^\ast}$
with an inverse effective mass
$(m^\ast)^{-1} \equiv \frac{\partial^2 H}{\partial p^2}=\cos{p} \, e^{-\eta}$.
The quasi-momentum dependence of the effective mass allows a rich
variety of dynamical regimes. Solitonic solutions with a positive 
non-linear parameter $\Lambda > 0$, for instance, are allowed
by a negative effective mass. 
A regime with a diverging effective mass $m^\ast \to \infty$ 
leads to a self-trapping of the wave packet.
  

In the following we will study the dynamical regimes of
Eq.~(\ref{DNLS}) in two particular cases. We first consider 
a tilted (washboard) periodic potential describing the vertical 
optical trap created in the Anderson-Kasevich experiment. 
In particular, we will show that non-linear effects tend to
destroy the Bloch oscillations, consistently with the experimental
results. The external potential for the tilted trap is given by 
Eq.~(\ref{potential_anderson_kasevich}) with
$\epsilon_n= \omega n$ where $\omega = \frac{mg \lambda / 2}{2 K}$.
We find $V = \omega \xi$ and $\dot{p} = - \omega$ \cite{nota2}. 

Then we consider the case of an horizontal 
array, in which gravitation only provides a constant energy 
shift, with $\dot{p}=0$.
We will classify four different regimes which
include discrete breathers and solitons.  

$Tilted \, \, \, trap$. It is well known that single
atoms in a tilted washboard potential oscillate among sites at the Bloch 
frequency. This regime is described by Eqs.~(\ref{p-xi}, \ref{alpha-delta}) 
with $\Lambda =0$ (corresponding to a negligible mean field
condensate interaction). This is, precisely, the regime 
investigated in \cite{anderson98} in which a coherent output was observed. 
Indeed a variational estimate gives $\Lambda \simeq 0.5 $ and $\omega \simeq 2$ (the scaled time is in units of $\hbar/2K=0.35 ms$).  
In this limit the equations of motion can be solved exactly giving 
$p(t) = - \omega t + p_0$, 
$\xi(t) = - A [\cos{p_0} - \cos{(\omega t - p_0)}]$,
$\gamma^2(t) = 4 A^2 \log{A \omega} \{ \cos{[2(\omega t -p_0)]} - 
\cos {2p_0} - 4 \sin{p_0} \sin{(\omega t - p_0)} - 4 \sin^2{p_0}\} + 
\gamma_0^2$ 
and $\delta(t)=- \frac{8 A \log{A \omega} \, 
[\sin{(\omega t - p_0)}+\sin{p_0}] }{\gamma^2(t)}$ where 
$A=-H_0/\omega \cos{p_0}$, $\delta(0)= \delta_0 = 0$ 
and  $H_0$ is the (conserved) initial energy.
In the inset of Fig. 1 we show that the Bloch 
oscillations described by the variational ansatz (solid line) are in 
excellent agreement with the full
numerical solution of the DNLSE (dashed line). 
The numerical average position is defined as $\sum_n n \mid \psi_n \mid^2$. 

The effect of non-linearity on the Bloch oscillations is dramatic.
This has been studied experimentally in   
\cite{anderson98} by increasing the density in each well, 
and observing a degradation in the interference pattern. 
With $\Lambda \ne 0$ we have: 
\begin{equation}
\label{eq-smorzata}
\ddot{\xi} + \frac{\Lambda \delta}{2\sqrt{\pi} \gamma} \dot{\xi} + \omega^2 \xi = \omega H_0 - \frac{\Lambda \omega}{2 \sqrt{\pi} \gamma}.
\end{equation}    
The Eq.~(\ref{eq-smorzata}) displays an effective 
damping term proportional to the velocity $\dot{\xi}$. 
We stress that the dynamics is fully Hamiltonian, and real 
dissipative processes 
are absent. For $t \to \infty$, 
$\gamma$ tends to a constant value $\gamma_{fin}$ and $\delta \sim  
\frac{2 \Lambda}{\sqrt{\pi} \gamma^3_{fin}} t$, 
so the term $\Lambda \delta$ has 
the correct positive sign (for large $t$). The apparent damping 
is the consequence of
a diverging effective mass 
of the wave packet $m^{\ast} \sim e^{\frac{\Lambda^2}
{2 \pi \gamma_{fin}^4} t^2 }$, 
which stops the Bloch oscillations. This effect manifests, in the numerical analysis of Eq.~(\ref{DNLS}), as a distortion of the on-site phases. 
In Fig.1 (solid line) we show the variational Bloch dynamics  
with a non-linear parameter $\Lambda = 10$ and initial values 
$\xi_0=0$, $p_0=0$ and $\delta_0=0$. The oscillation 
roughly decreases as $\xi(t) \sim - A (1 - e^{-\frac{\Lambda^2 t^2}
{2 \pi \gamma_{fin}^4}} \cos{\omega t} )$.
The dashed line shows the full numerical solution of the
DNLSE, in good agreement with the analytical result (solid line). 
The discrepancy at 
$t > 10$ is due to the breaking of the gaussian wave packet
in the numerical simulation.

$Untilted \, \, \, trap$. 
The momentum $p(t)=p_0$ is conserved.
We note that the equations of motion are invariant with respect to the replacement
$\Lambda \to - \Lambda$, $p_0 \to p_0 + \pi$ and $t \to -t$. 
In Fig. 2 we report the dynamical phase diagram as a function of 
$\Lambda$ versus $\cos{p}$, for $\Lambda > 0$ and 
with initial values $\xi_0=0$ and $\delta_0=0$. 
This phase diagram has been calculated analytically from 
Eqs.~(\ref{p-xi}-\ref{Hamiltonian-DNLS}) and checked numerically. 
In the region $\cos{p}>0$ 
there are two distinct regimes.
When $H_0  > 0$, $\gamma(t) < \gamma_{max}$: 
this is the self-trapped regime in which the boson wave
packet remains localized around few sites. 
The self-localization is a genuine non-linear effect, characterized by a
diverging effective mass.
In particular, the self-trapped wave packet cannot translate 
along the array: this is a major difference with respect to
soliton-like solutions. 
The limit values for $t \to \infty$  
are $\gamma \to \frac{\Lambda}{2\sqrt{\pi} H_0}$,
$\delta \to \infty$ and $\dot{\xi} \to 0$. 
We note that a non-linear
self-trapping occurs also in a two-site
problem \cite{smerzi97}.

A diffusive regime occurs when $-\cos{p_0}<H_0 \leq 0$. In this case 
$\gamma(t \to \infty) \to \infty$ and 
$\dot{\xi} \approx -H_0/\tan{p_0} = const$. 
The transition between the two regimes occurs
at $\Lambda_c=2 \sqrt{\pi} \, \gamma_0 \, \cos{p_0} \, e^{-1/2\gamma_0^2}$. 
With $\Lambda>\Lambda_c$, the ratio between the initial value of the width $\gamma_0$ and the 
limit width $\gamma_{max}(t \to \infty)$ is given by 
\begin{equation}
\label{alpha_max}
\frac{\gamma_0}{\gamma_{max}}= \frac{\Lambda - \Lambda_c}{\Lambda}.
\end{equation} 
In Fig. 3 we plot $\gamma_0/\gamma_{max}$ vs. $\Lambda/\Lambda_c$ 
while in the inset we report the 
variational and numerical values of the width of the density 
vs. $\Lambda/\Lambda_c$ after a time $\tau=10$ (scaled units). 
The discrepancy at large
$\Lambda / \Lambda_c$ is due to a slight deviation of the numerical density
profile from a gaussian shape.
We checked the stability of self-trapping, also considering 
different initial forms of the wave packet. 

In the region $\cos{p}<0$ 
the self-trapping condition is given by $H_0>\mid \cos{p_0} \mid$: in this case 
$\Lambda_c=2 \sqrt{\pi} \, \gamma_0 \mid \cos{p_0} \mid \, (1-e^{-1/2\gamma^2_0})$.
A soliton solution can be determined by imposing 
$\dot{\gamma}=\dot{\delta}=0$. We find 
$\Lambda_{sol} = 2 \sqrt{\pi} \frac{\mid \cos{p_0} \mid} 
{\gamma_0} \, e^{-1/2\gamma^2_0}$. 
For $\Lambda=\Lambda_{sol}$ the center of the wave packet moves with 
a constant velocity $\dot{\xi}$ and its width remains 
constant in time. 
For $\Lambda_c<\Lambda<\Lambda_{sol}$, $\xi \to \infty$ while 
$\gamma(t)$ oscillates, 
corresponding to a breather solution. 
The numerical solution of Eq.~(\ref{DNLS}) confirms 
the variational predictions. In Fig. 4(a) 
we report the numerical density profile for $\Lambda$ 
in the breather region with $p_0=3\pi/4$; in Fig. 4(b) the 
numerical density profile is drawn for the same $\Lambda$ and $p_0=\pi/4$: 
the change of the sign of $\cos{p}$ gives a breather solution 
in the first case, and the spreading of the density 
in the second one. In Figs. 4(c) and 4(d) we show that for 
$\Lambda=\Lambda_{sol}$ the soliton after hitting a wall
rebounds and regains its shape. 

In conclusion, we have studied the dynamics of a dilute Bose condensate
trapped in a periodic potential.
We have analyzed the Anderson-Kasevich experiment, pointing out that
mean field effects lead to a coherent destruction of the interwell
Bloch oscillations. 
We have shown that intrinsically localized excitations, 
like discrete breathers and solitons, can exist also with 
a repulsive interatomic potential by studying analytically 
their dynamical phase diagram.

Discussions with S.R. Shenoy and P. Sodano are acknowledged. This work has been supported by the Cofinanziamento MURST.

\begin{figure}[h]
\caption{Coherent destruction of Bloch oscillations: numerical (dashed line)
 and variational (solid line) average position of the density 
in the tilted trap ($\omega=2$, $\Lambda=10$ and initial $p_0=0$, 
$\delta_0=0$, 
$\gamma_0=10$). Inset: Bloch oscillations for $\Lambda=0$.}    
\end{figure}

\begin{figure}[h]
\caption{Dynamical phase diagram in the untilted trap: $\Lambda$ vs. $\cos{p}$, for $\Lambda > 0$ and with initial $\xi_0=0$ and $\delta_0=0$.}
\end{figure}

\begin{figure}[h]
\caption{Universal curve for $\gamma_0/\gamma_{max}$ vs. 
$\Lambda/\Lambda_c$ as in Eq.~(\ref{alpha_max}). The inset shows the 
variational (solid line) and numerical (circles) width
of the density vs. $\Lambda/\Lambda_c$ after the time $\tau=10$, 
scaled units (initial values: $p_0=0$, $\delta_0=0$ and $\gamma_0=7$).}
\end{figure}

\begin{figure}[h]
\caption{(a) Numerical density profiles calculated at different times
($t = 0, 50, 100,\ldots,400$).
The value of $\Lambda$ is inside 
the breather region of the dynamical phase diagram
($p_0=3 \pi /4,~\gamma_0 = 10,~\Lambda = 0.24$); 
(b) the same parameters as in (a), but for $p_0=\pi /4$; 
(c) width (dotted line) and average position (solid line) 
calculated numerically for $\Lambda=\Lambda_{sol}$ 
and $p_0=3 \pi /4$ in a finite array of $73$ sites 
(the soliton hits a wall); 
(d) numerical density profile for $t=0,50,100,150,200$ 
for the soliton described in (c).}
\end{figure}


\begin{thebibliography}{10}
\bibitem{nota1}
With $\gamma$ not too small ($\gamma > 1$) 
we can replace the sums over $n$ in the Lagrangian with integrals:
$\sum\limits_{n=-\infty}^{\infty} e^{-\frac{(n-\xi)^2}{\gamma^2}} 
\Bigg / \int\limits_{-\infty}^{\infty} 
e^{-\frac{(n-\xi)^2}{\gamma^2}} dn = 1+e^{-\pi^2 \gamma^2} 
\cos{(2\pi\xi)}+O(e^{-4 \pi^2 \gamma^2})$.  
In this limit the normalization factor becomes $k=\sqrt{2/\pi \gamma^2}$.
\bibitem{nota2} 
We note that for a quadratic potential 
$\epsilon_n= \Omega n^2$ we would have 
$V(\gamma, \xi) = \Omega (\xi^2 + \gamma^2 / 4)$ 
and $\dot{p} = - 2 \Omega \xi$.
\end{thebibliography}
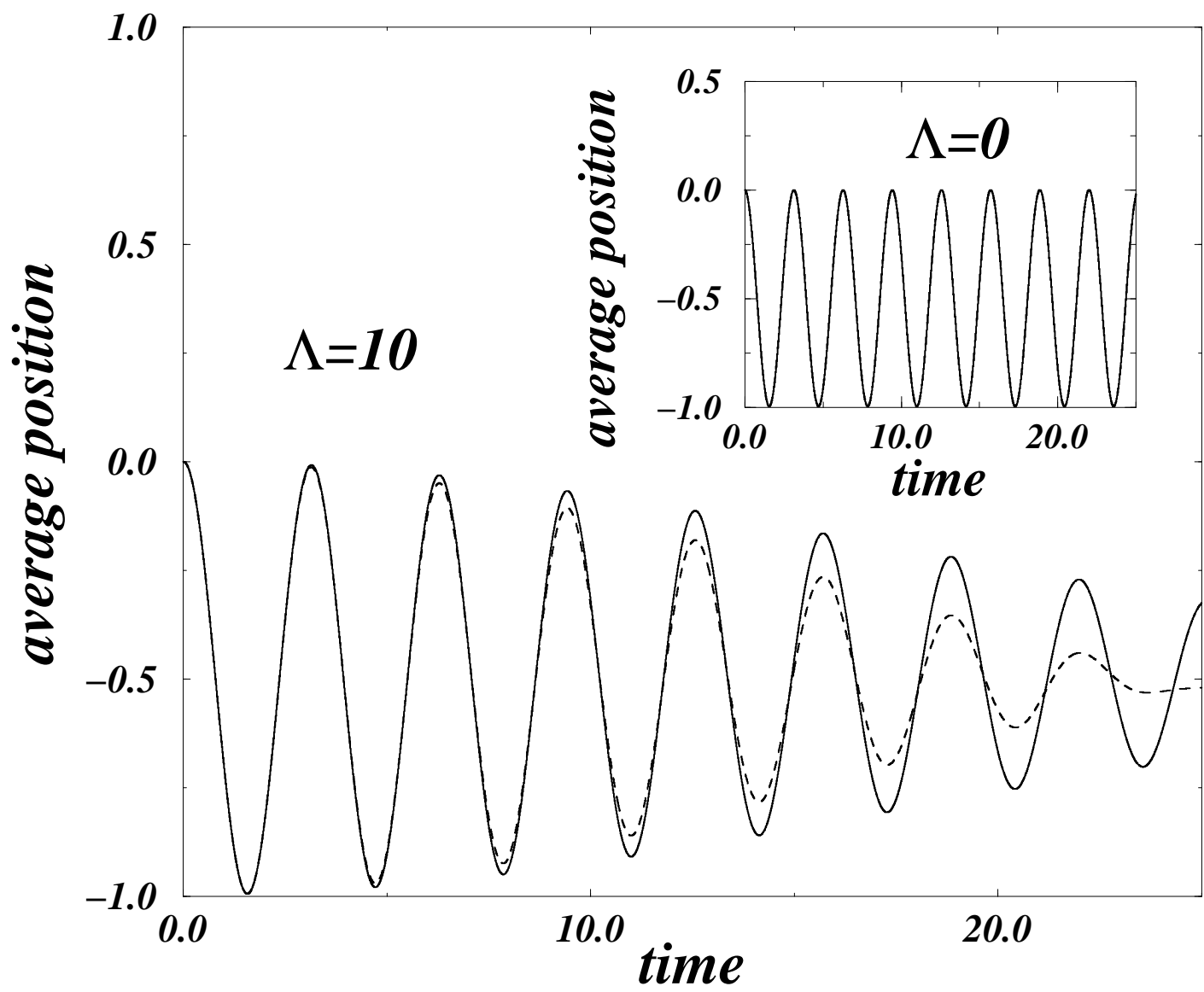
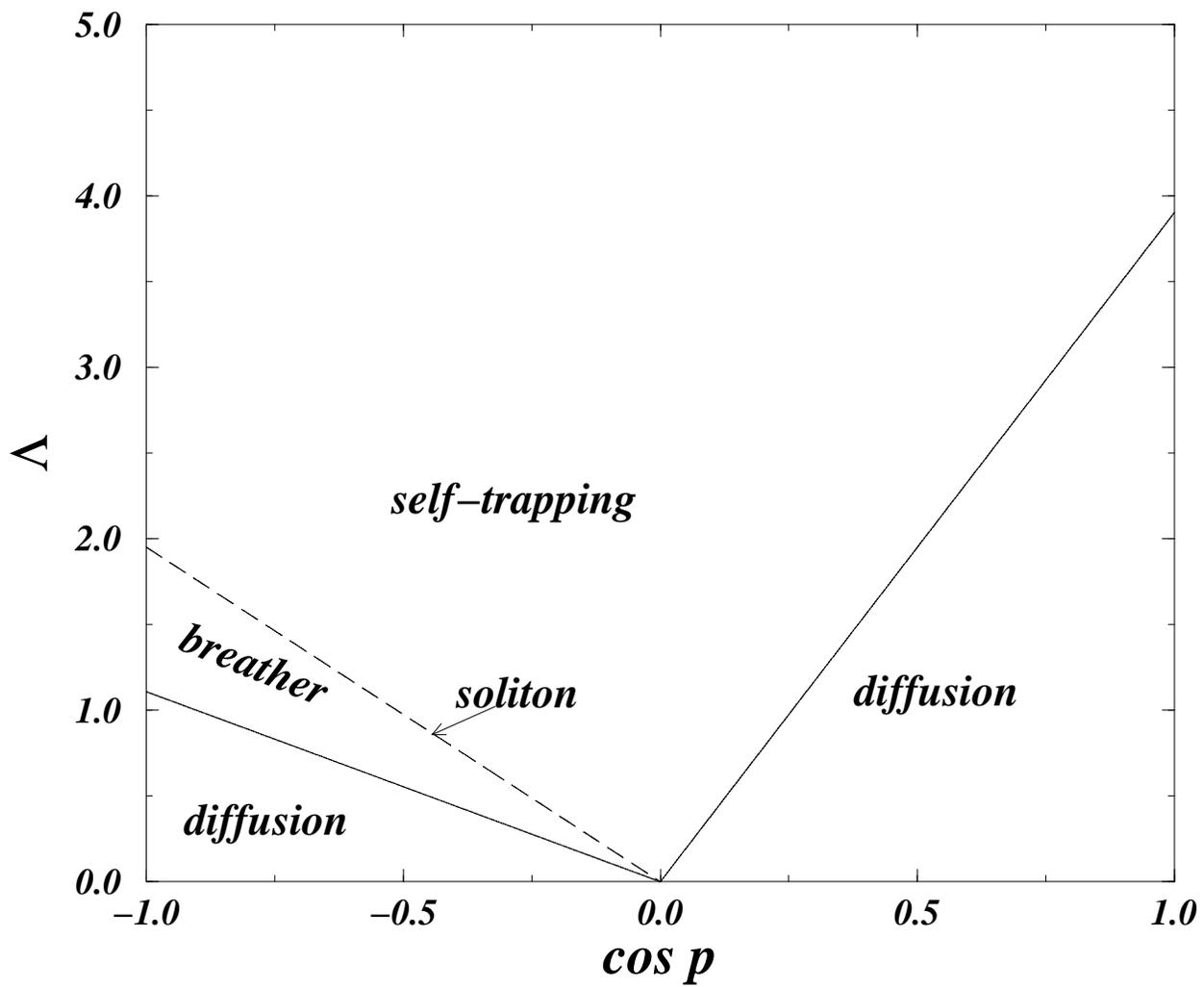
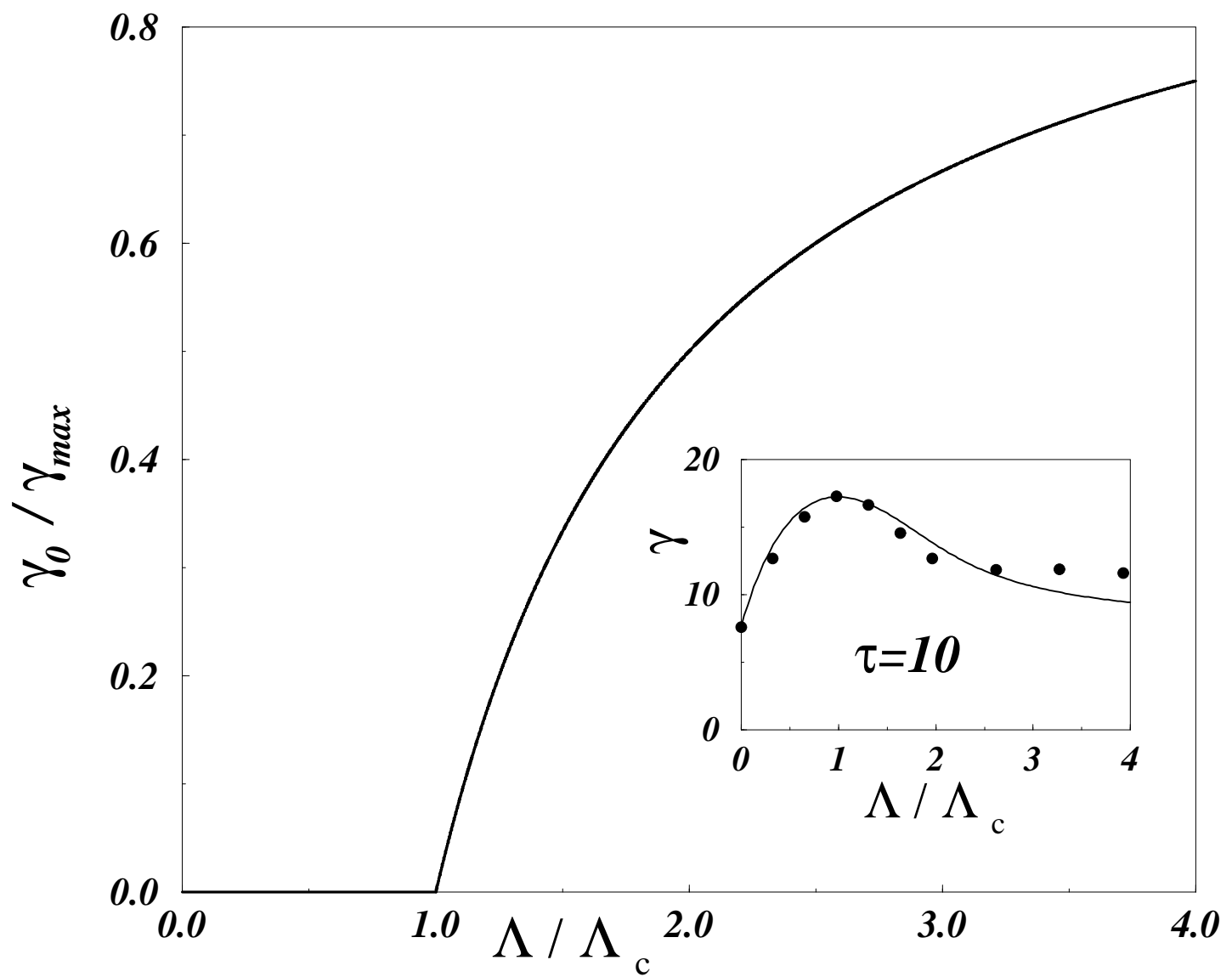
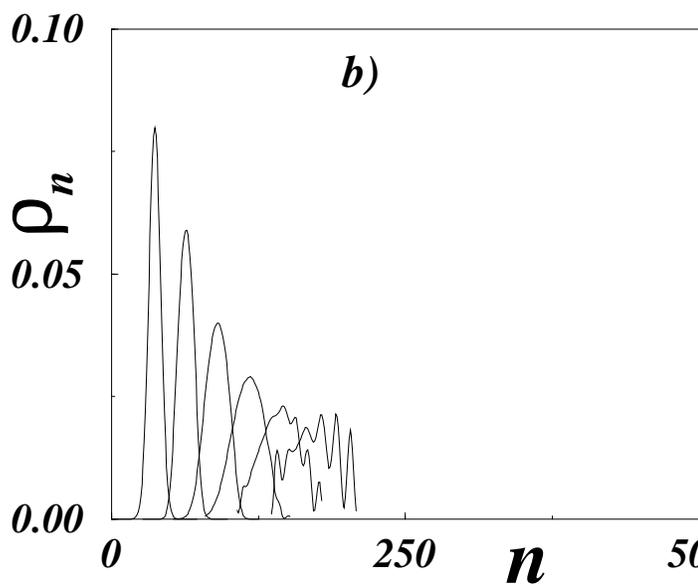
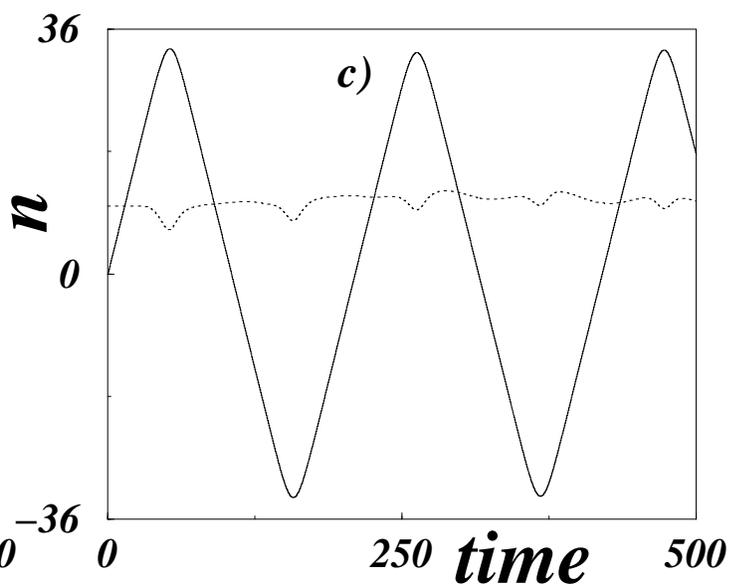
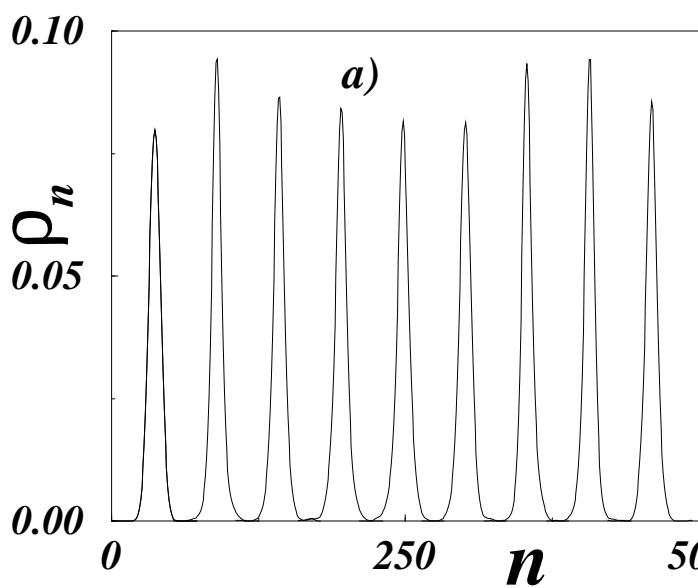
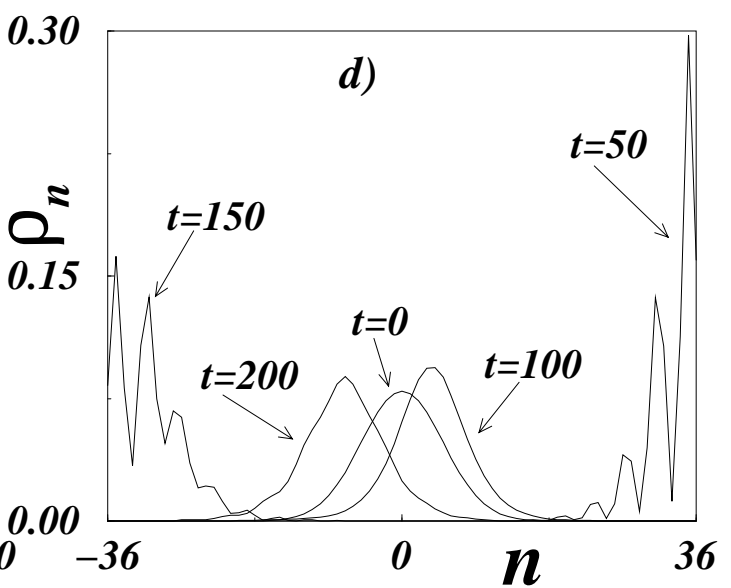
\end{document}